\documentclass[manuscript, review, screen]{acmart}
\usepackage{url}

\acmJournal{JDIQ}
\acmVolume{0}
\acmNumber{0}
\acmArticle{0}
\acmYear{2018}
\acmMonth{0}

\setcopyright{acmcopyright}

\acmDOI{0000001.0000001}



\title{Transparency, Fairness, Data Protection, Neutrality: Data Management Challenges in the Face of New Regulation}

\author{Serge Abiteboul}
\affiliation{%
  \institution{Inria \& Ecole Normale Sup\'erieure}
  \country{France}}
\email{Serge.Abiteboul@inria.fr}
\author{Julia Stoyanovich}
\affiliation{%
  \institution{New York University}
  \country{USA}
}
\email{stoyanovich@nyu.edu}

\begin{document}

\begin{abstract}
The data revolution continues to transform every sector of science, industry and government. Due to the incredible impact of data-driven technology on society, we are becoming increasingly aware of the imperative to use data and algorithms responsibly --- in accordance with laws and ethical norms.  In this article we discuss three recent regulatory frameworks: the European Union's General Data Protection Regulation (GDPR), the New York City Automated Decisions Systems (ADS) Law, and the Net Neutrality principle, that aim to protect the rights of individuals who are impacted by data collection and analysis.  These frameworks are prominent examples of a global trend: Governments are starting to recognize the need to regulate data-driven algorithmic technology. 

Our goal in this paper is to bring these regulatory frameworks to the attention of the data management community, and to underscore the technical challenges they raise and which we, as a community, are well-equipped to address.  The main take-away of this article is that legal and ethical norms cannot be incorporated into data-driven systems as an afterthought.  Rather, we must think in terms of responsibility by design, viewing it as a systems requirement.  
\end{abstract}

\maketitle

\section{Introduction}
\label{sec:intro}

The data revolution continues to transform every sector of science, industry and government. Due to the incredible impact of data-driven technology on society, we are becoming increasingly aware of the imperative to use data and algorithms {\em responsibly} --- in accordance with laws and ethical norms. 
The goal of this article is to underscore the technical challenges raised by recent legal and regulatory frameworks, which the data management community is well-equipped to address.

We discuss three recent frameworks: the European Union's General Data Protection Regulation (GDPR)~\cite{gdpr}, the New York City Automated Decisions Systems (ADS) Law~\cite{Vacca}, and the Net Neutrality principle.  These frameworks are prominent examples of a global trend: Governments are starting to recognize the need to regulate data-driven algorithmic technology.  The GDPR and the NYC ADS Law aim to protect the rights of individuals who are impacted by data collection and analysis, while the Net Neutrality principle ensures that  services are being treated equitably.  Yet, despite the focus on organizations, rights of individuals also figure prominently in the neutrality debate: One of the imperatives is that individuals should be able to enjoy freedom of choice and expression on-line.  We will give some legal context on neutrality by discussing the EU Regulation 2015/2120~\cite{EU2015_2120}, the Indian Net Neutrality Regulatory Framework~\cite{india}, and the ongoing regulatory debate on Net Neutrality in the US. 

Our goal in this paper is to bring these regulatory frameworks to the attention of the data management community. The main take-away of this article is that legal norms cannot be incorporated into data-driven systems as an afterthought.  Rather, we must think in terms of {\em responsibility by design}, viewing it as a systems requirement.  

\subsection{The General Data Protection Regulation}
\label{sec:intro:gdpr}

The European Union recently enacted a sweeping regulatory framework known as the General Data Protection Regulation, or the GDPR~\cite{gdpr}.  The regulation was adopted in April 2016, and became enforceable about two years later, on May 25, 2018.  The GDPR aims to protect the rights and freedoms of natural persons with regard to how their personal data is processed, moved and exchanged (Article 1).  The GDPR is broad in scope, and applies to ``the processing of personal data wholly or partly by automated means'' (Article 2), both in the private sector and in the public sector.  Personal data is broadly construed, and refers to any information relating to an identified or identifiable natural person, called the {\em data subject} (Article 4).  In this article we focus on the following salient points of the regulation:

\begin{itemize}
\item lawful processing of data is predicated on the data subject's informed consent, stating whether their personal data can be used, and for what purpose (Articles 6, 7);
\item the data subject has a right to correct any errors in their data (``right to rectification'', Article 16), to withdraw their data from the system (``right to erasure'', Article 17), and to move data from one data processor to another (``right to portability'', Article 20);
\item the data subject has the right to be informed about the collection and use of their data.~\footnote{\url{https://gdpr-info.eu/issues/right-to-be-informed/}}
\end{itemize}

The primary focus of the GDPR is on protecting the rights of data subjects, by giving them insight into, and control over, the collection and processing of their personal data.  Providing insight, in response to the ``right to be informed'', requires technical methods for algorithmic and data transparency, which we will discuss in Section~\ref{sec:transp}.  
We will also discuss the challenges inherent in giving individuals an ability to erase or move their data in Section~\ref{sec:moveremove}.

\subsection{The New York City Algorithmic Decision Systems Law}
\label{sec:intro:ads}

New York City recently passed a law~\cite{Vacca} requiring that a task force be put in place to survey the current use of ``automated decision systems'' (ADS), defined as ``computerized implementations of algorithms, including those derived from machine learning or other data processing or artificial intelligence techniques, which are used to make or assist in making decisions,'' in City agencies.  The task force is working to develop a set of recommendations for enacting algorithmic transparency by the agencies, and will propose procedures for:  

\begin{itemize}
\item requesting and receiving an explanation of an algorithmic decision affecting an individual (Section 3 (b)); 
\item interrogating automated decision systems for bias and discrimination against members of legally protected groups, and addressing instances in which a person is harmed based on membership in such groups (Sections 3 (c) and (d));
\item assessing how automated decision systems function and are used, and archiving the systems together with the data they use (Sections 3 (e) and (f)).
\end{itemize}

In contrast to the GDPR, which is very broad in scope, the NYC ADS Law only regulates City agencies in their use of algorithms and data, and does not directly apply to private companies.  However, because government agencies often procure systems and components from industry partners, the Law will likely impact industry practices.  Further, while New York is the first US city to pass a law of this kind, we expect other US municipalities to follow with similar legal frameworks or recommendations in the near future.  

The primary focus of the NYC ADS Law is on algorithmic transparency, which, in turn, cannot be achieved without data transparency~\cite{follow}.  As we discussed in Section~\ref{sec:intro:gdpr}, transparency is also an implicit requirement of the GDPR, stemming from the ``right to be informed''.  We will discuss the role that the data management community can play in enabling data transparency in Section~\ref{sec:transp}.  

The NYC ADS Law further requires fair and equitable treatment of individuals, mandating that ADS safeguard against bias and discrimination, and provide transparency in this regard.  We will discuss fairness in Section~\ref{sec:fair}, and will propose some research directions for the data management community that are complementary to the rich and rapidly expanding body of work on fairness in machine learning.  

\subsection{The Net Neutrality Principle}
\label{sec:intro:net}

Net Neutrality is the principle that Internet Service Providers (ISPs) should not discriminate or charge differently based on the message source (the content provider), its destination (the user), or its content. The concept was articulated by Tim Wu in 2003~\cite{Wu}.

According to Net Neutrality, an ISP cannot block or  throttle video streams from YouTube (negative discrimination), or enable free access to Facebook out of package (a kind of positive discrimination). A September 2018 report from Northeastern University and the University of Massachusetts, Amherst, found that US telecommunications companies are indeed slowing internet traffic to and from those two sites in particular, along with other popular apps~\cite{wehe,Kharif}. Of course, there are limits to the non-discrimination, such as blocking pornographic material for young Internet users, filtering hate speech in some countries, or guaranteeing quality for emergency services. 

In the European Union, Net Neutrality is guaranteed by EU Regulation 2015/2120~\cite{EU2015_2120}, although different countries may interpret the regulation differently.  For example, some forms of zero-rating, the practice of providing Internet access without financial cost as a means of positive discrimination, are legal in some EU countries but not in others. Since 2018, India has perhaps the world's strongest Net Neutrality rules~\cite{india}. In general, more and more countries are adopting Net Neutrality regulations, with a notable exception. In the United States, the Federal Communications Commission (FCC) issued its Open Internet Order in 2015, reclassifying Internet access --- previously classified as an information service --- as a common carrier telecommunications service, thereby enforcing some form of Net Neutrality. However, in 2017, under the chairmanship of Ajit Pai, the FCC officially repealed Net Neutrality rules.

\section{Algorithmic and Data Transparency}
\label{sec:transp}

ProPublica's story on ``machine bias'' in an algorithm used for sentencing defendants~\cite{propublica} amplified calls to make algorithms more transparent and accountable~\cite{Kroll2017}.  Transparency and accountability are intrinsically linked with trust, and are of particular importance when algorithmic systems are integrated into government processes,  assisting humans in their decision-making tasks, and sometimes even replacing humans. Transparency of government is a core democratic value, which compels us to develop technological solutions that both increase government efficiency and can be made transparent to the public.

A narrow interpretation of algorithmic transparency requires that the source code of a system be made publicly available. This is a significant step towards transparency (as long as the posted code is readable, well-documented and complete), but it is rarely sufficient.  One of the reasons for this, of particular relevance to the data management community, is that meaningful transparency of algorithmic processes cannot be achieved without transparency of data~\cite{follow}. 

What is data transparency, and how can we achieve it? One immediate interpretation of this term in the context of predictive analytics includes ``making the training and validation datasets publicly available.'' However, while data should be made open whenever possible, much of it is sensitive and cannot be shared directly. That is, data transparency is in tension with the privacy of individuals who are included in the dataset.  In light of this, we may adopt the following alternative interpretation of data transparency:  In addition to releasing training and validation datasets whenever possible, vendors should make publicly available summaries of relevant statistical properties of the datasets that can aid in interpreting the decisions made using this data, while applying state-of-the-art methods to preserve the privacy of individuals (such as differential privacy~\cite{DBLP:journals/fttcs/DworkR14}).  When appropriate, privacy-preserving synthetic datasets can be released in lieu of real datasets to expose certain features of the data~\cite{DBLP:conf/ssdbm/PingSH17}.

An important aspect of data transparency is interpretability --- surfacing the statistical properties of a dataset, the methodology that was used to produce it, and, ultimately, substantiating its ``fitness for use'' in the context of a specific automated decision system or task.  This consideration of a specific use is particularly important because datasets are increasingly used outside the original context for which they were intended.  The data management community can begin addressing these challenges by building on the significant body of work on data profiling (see~\cite{DBLP:conf/sigmod/AbedjanGN17} for a recent tutorial), with an eye on the new legal requirements.    

Interpretability rests on making explicit the interactions between the program and the data on which it acts.  This property is important both when an automated decision system is interrogated for systematic bias and discrimination, and when it is asked to explain an algorithmic decision that affects an individual.  For example, suppose that a system scores and ranks individuals for access to a service. If an individual enters her data and receives the result --- say, a score of 42 --- this number alone provides no information about why she was scored in this way, how she compares to others, and what she can do to potentially improve her outcome.   A prominent example of a system of this kind, which is both opaque and extremely impactful, is the FICO credit scoring system in the US~\cite{CitronP14}. 

The data management research community is well-positioned to contribute to developing new  methods for interpretability.  These new contributions can naturally build on a rich body of work on data provenance (see~\cite{DBLP:journals/vldb/HerschelDL17} for a recent survey), on  recent work on explaining classifiers~\cite{DBLP:conf/kdd/Ribeiro0G16} and auditing black box models using causal framework~\cite{DBLP:conf/sp/DattaSZ16}, and on automatically generating ``nutritional labels'' for data and models~\cite{DBLP:conf/sigmod/YangSAHJM18}.

\section{Fairness}
\label{sec:fair}

We can all agree that algorithmic decision-making should be fair, even if we do not agree on the definition of fairness. But isn't this about algorithm design? Why is this a data problem?  Indeed, the machine learning and data mining research communities are actively working on methods for enabling fairness of specific algorithms and their outputs, with a particular focus on classification problems (see, for example,~\cite{DBLP:conf/innovations/DworkHPRZ12,DBLP:conf/kdd/FeldmanFMSV15,DBLP:journals/corr/FriedlerSV16,DBLP:journals/tkde/HajianD13,DBLP:journals/kais/KamiranZC13,DBLP:conf/innovations/KleinbergMR17,DBLP:journals/ker/RomeiR14} and proceedings of the recently established ACM Conference on Fairness, Accountability, and 
Transparency (ACM FAT*)~\footnote{\url{https://www.fatconference.org/}}).  While important, these approaches focus solely on the final step in the data science lifecycle, and are thus limited by the assumption that input datasets are clean and reliable.

Data-driven algorithmic decision making usually requires multiple pre-processing stages to address messy input and render it ready for analysis~\cite{DBLP:journals/cacm/JagadishGLPPRS14}. This pre-processing, which includes data cleaning, integration, querying and ranking, is often the source of algorithmic bias~\cite{Kirkpatrick:2017:AD:3042068.3022181,DBLP:conf/ssdbm/StoyanovichHAMS17}, and so reasoning about sources of bias, and mitigating unfairness upstream from the final step of data analysis, is potentially more impactful.

For example, much research goes into ensuring statistical parity --- a requirement that the demographics of those receiving a particular outcome, (e.g., a positive or negative classification), are identical to the demographics of the population as a whole.  Suppose that the input to a binary classifier contains 900 men and 100 women, but that it is known that women represent 50\% of the over-all population, and so achieving statistical parity amounts to enforcing a 50-50 gender balance among the positively classified individuals. That is, all else being equal, a woman in the input to the classifier is far more likely to receive a positive classification than a man. An alternative is to observe the following: If the input to the classifier was produced by a SQL query, and if relaxing the query would make the input more balanced (e.g., 1000 men and 500 women), then a more effective way to mitigate the lack of statistical parity in the output of the classifier is to relax the query upstream.

It is easy to construct additional examples that show how bias may be introduced during data cleaning, data integration, querying, and ranking --- upstream from the final stage of data analysis. Therefore, it is meaningful to detect and mitigate these effects in the data lifecycle stages in which they occur. (See~\cite{shira} for a discussion of the definitions of ``bias'', and of the corresponding assumptions made when defining fairness measures.)

Members of the data management community who are interested in this topic may consider a growing body of work on impossibility results, which show that different notions of fairness cannot be enforced simultaneously, and so require explicit trade-offs~\cite{DBLP:journals/corr/Chouldechova17,DBLP:journals/corr/FriedlerSV16,DBLP:conf/innovations/KleinbergMR17}. These are not negative results per se, nor are they surprising. Fairness is a subjective, context-dependent and highly politicized concept; a global consensus on what is fair is unlikely to emerge, in the context of algorithmic decision making or otherwise.  Think, for example, of the decade-long debate about the interplay between ``disparate treatment'' and ``disparate impact'', for which recent examples include by Ricci v. De Stefano~\footnote{\url{https://en.wikipedia.org/wiki/Ricci_v._DeStefano}} and the ongoing lawsuit regarding the use of race in Harvard University admissions~\footnote{\url{https://www.nytimes.com/2018/10/13/us/harvard-affirmative-action-asian-students.html}}.  That being said, a productive way to move forward in the data science context is to develop methods that can be instrumented with different alternative fairness notions, and that can support principled and transparent trade-offs between these notions.

\section{Moving and Removing Personal Data}
\label{sec:moveremove}

\subsection{The Right to Be Forgotten}
\label{sec:forget}

The right to be forgotten is originally motivated by the desire of individuals to not be perpetually stigmatized by something they did in the past. Under pressure from despicable social phenomena such as revenge porn, it was turned recently into laws in 2006 in Argentina, and since then in the European Union, as part of the GDPR.  In particular, Article 17 of the GDPR states that data subjects have the right to request erasure of their personal data, and that they can do so for a large number of reasons.

The passing of this law primarily resulted in a high number of requests to search engines to dereference web pages. This turned out to be controversial for a number of reasons, including also that the dereferencing by Google is very opaque, and that this company in effect acquired, against its own will, a questionable power to adjudicate.  Furthermore, as is advocated by Wikimedia among others, the right to be forgotten sometimes conflicts with other rights such as the public's right to information.

In addition to search engines, the right to be forgotten affects companies that keep personal data.  A prominent example is Facebook, where for many years it was impossible to delete data that pertains to a user's account.  A user may close an account, then reopen it some time later and find all her data as it was originally. It is now possible to request the deletion of all data pertaining to an account from Facebook, however, the user has no proof that the deletion indeed occurred. 

An important technical issue, of clear relevance to the data management community, is that of deletion of information in systems that are typically meant to accumulate data.  This deletion must be both permanent and deep, in the sense that its effects must propagate through data dependencies. To start, it is difficult to guarantee that all copies of every piece of deleted data have actually been deleted.  Further, when some data is deleted, the remaining database may become inconsistent, and may, for example, include dangling pointers. Additionally, production systems typically do not include a strong provenance mechanism, and so they have no means of tracking the use of an arbitrary data item (one to be deleted), and reasoning about the dependencies on that data item in derived data products.  

Although much attention of the data management community has over the years been devoted to tracking and reasoning about provenance, primarily in relational contexts and in workflows (see~\cite{DBLP:journals/vldb/HerschelDL17} for a recent survey), there is still important work to be done on making these methods both practically feasible, and sufficiently general to accommodate the current legal requirements.  An important direction that is, to the best of our knowledge, still unexplored, concerns ascertaining the effects of a deletion on downstream processes that are not purely relational, but include other kinds of data analysis tasks, like data mining or predictive analytics.

Requests for deletion may also conflict with other laws such as requirements to keep certain transaction data for some period of time, or with requirements for fault tolerance and recoverability.  Should the deleted pieces of data also be erased from caches and backups? Requesting this functionality gives immediate nightmares to systems engineers in charge of a production data management system, with millions of lines of code and terabytes of legacy data. The likely answer is: ``this cannot be done; the only solution I see is redeveloping the system from scratch with right-to-be-forgotten-by-design.''  Understanding the impact of deletion requests on our ability to offer guarantees on system resilience and performance, and developing appropriate primitives and protocols for practical use, is another call to action for the data management community.

\subsection{Interoperability and Portability}
\label{sec:portability}

Article 20 of the GDPR, ``Right to data portability'', stipulates a data subject's right to receive her personal data from a vendor, and to transfer her data to another vendor. The main goals of this provision are both to keep the data subject informed about what data a vendor has about her, and to prevent vendor lock-in. This enables a user who is unhappy with a service to leave for a competing service that best serves her needs, without having to reconstruct her entire data history. This also allows a user to select applications of her choice and have them cooperate, to her best advantage, even if they come from different vendors.

In response to data portability regulation, and to users' concerns, Google, Twitter, Microsoft, and Facebook teamed up in the Data Transfer Project that aims to facilitate content transfer between applications. Of course, it is not an easy task for a company to provide a service that facilitates the departure of its customers. This is why, in spite of commendable behavior of companies that engage in the Data Transfer Project, it is the role of regulators to impose data portability and interoperability requirements. 

Interoperability of database applications is an old topic. But one can imagine an unlimited number of possibilities, such as having a Whatsapp call talk to a Skype one. And it certainly acquires a different flavor when we consider interoperating applications with billions of users and millions of transactions per second.

For data portability, it should be noted that the devil is in the detail. The export format should be stable and structured to facilitate reuse. Also, which data can be exported is an issue. Obviously, it includes all data that the user volunteered to the service. But should it also include data the vendor gathered from the behavior of the user (e.g., the time the user is waking up in the morning)? Should it include data the service inferred (e.g., what is the home address of the user, her job address)? 

Another issue with portability is the target system. A user may want to port her photos from Service A to Service B. The issue is then for Service B to be able to incorporate as much data as possible from Service A. Now, the user may want to integrate her photos in a personal information system \cite{AK}. Such a system must be able to integrate information from a large panel of domains. This brings us to the fields of data integration \cite{lenzerini} and knowledge representation. 

\section{Neutrality}
\label{sec:neutral}

As already mentioned, Net Neutrality is now legally required in some countries.  Yet, detecting Net Neutrality violations to enforce the law is not an easy task. Indeed, simply measuring the performance of Internet communications is not easy: measurement results may depend on the location of the source, of the target, of the context (other applications competing for the same bandwidth), and on other factors. Indeed, different measures provided for network traffic typically diverge. The evaluation of Net Neutrality relying on such hard-to-obtain measures is a challenging research topic~\cite{wehe}, which is primarily of interest to the networks and Internet measurement communities, and less so to data management. 

But beyond Net Neutrality, new forms of neutrality are emerging such as device neutrality (is my smart-phone blocking certain apps and favoring others?), and platform neutrality (is this particular web service providing neutral recommendation?). For instance, app stores like Google Play and the Apple App Store, tend to refuse to reference certain services, perhaps because they are competing with the company's own services. Research is needed to be able to verify these new facets of neutrality. In particular, it is not easy to check whether a recommendation engine like Google search or Booking is enforcing only transparent editorial policies, and whether, other than that, their results are comprehensive, impartial and based solely on relevance. For example, it has been observed that search engines tend to favor some ``friendly'' services over competitors~\footnote{\url{https://en.m.wikipedia.org/wiki/European_Union_vs._Google}}.

\section{Take-aways}
\label{sec:conc}

In this article, we discussed several recent regulatory frameworks that aim to protect the rights of individuals, to ensure equitable treatment of services, and to bring transparency to data-driven algorithmic processes in industry and in government.  Our goal was to bring these regulatory frameworks to the attention of the data management community, and to underscore the technical challenges they raise and which we, as a community, are well-equipped to address.  

An important take-away of this article is that legal norms cannot be incorporated into data-driven systems as an afterthought.  Rather, we must think in terms of {\em responsibility by design}, viewing it as a systems requirement.  

We also stress that enacting algorithmic and data transparency, fairness, data protection, and neutrality will require a significant cultural shift.  In making this shift, we must accept that the objectives of ``efficiency'', ``accuracy'' and ``utility'' cannot be the primary goal, but that they must be balanced with equitable treatment of members of historically disadvantaged groups, and with accountability and transparency to individuals affected by algorithmic decisions and to the general public.

In this article we focused on explicit regulation of industry stakeholders by government entities (in the case of the GDPR and the Net Neutrality laws), and on government oversight (in the case of the NYC ADS law).  Another implicit regulatory mechanism can be achieved by empowering users and user associations, by providing them with data literacy education and with precise information on how different products and services work. Better educated users can choose better solutions, including more effective ways to protect their private data.  Such users can also more easily understand explanations provided to them by an algorithmic system. User associations can help individuals make informed choices, and support them via class actions lawsuits in the case of disputes.

\begin{acks}
This work was supported in part by \grantsponsor{001}{National Science Foundation (NSF)}{} Grant No. \grantnum{001}{1741047}, and by \grantsponsor{}{Agence Nationale de la Recherche (ANR)}{} Grant Headwork.

\end{acks}

\bibliographystyle{ACM-Reference-Format}
\bibliography{julia,bigdata}

\end{document}